\documentclass[11pt]{article}

\usepackage{a4wide}
\usepackage{amsmath,amssymb,amsthm,amscd,framed}
\usepackage[pdftex]{graphicx}

\usepackage[all]{xy}

\newcommand{\bb}[1]{\begin{equation}\label{#1}}
\newcommand{\ee}{\end{equation}}
\newcommand{\vc}[1]{{\bf #1 }}

\newcommand{\const}{\operatorname{const}}

\newcommand{\tr}{\operatorname{tr}}

\newcommand{\Cov}{\operatorname{Cov}}

\newcommand{\Var}{\operatorname{Var}}

\def\identy{{\mathsurround0pt\mathchoice{\textidenty}{\textidenty}{\scptidenty}{\scptidenty}}}
\def\scptidenty{\setbox0\hbox{$\scriptstyle1$}\bothidenty}
\def\textidenty{\setbox0\hbox{$1$}\bothidenty}
\def\bothidenty{\rlap{\hbox to.97\wd0{\hss\vrule height.06\ht0 width.82\wd0}}
 \copy0\rlap{\kern-.36\wd0\vrule height1.05\ht0 width.05\ht0}\kern.14\wd0}
\newtheorem{theorem}{Theorem}[section]

\theoremstyle{definition}

\newcommand{\bt}{\pmb\theta}

\begin{document}
\title{A Simulated Annealing Approach to Approximate Bayes Computations}

\author{Carlo Albert\footnote{Eawag, aquatic research, 8600 D\"ubendorf, Switzerland.},
Hans R. K\"unsch\footnote{Seminar f\"ur Statistik, ETH Z\"urich, 8092 Z\"urich, Switzerland.} and
Andreas Scheidegger\footnotemark[1]}
\maketitle

\begin{abstract}
Approximate Bayes Computations (ABC) are used for parameter inference when the likelihood function of the model is expensive to evaluate but relatively cheap to sample from.
In particle ABC, an ensemble of particles in the product space of model outputs and parameters is propagated in such a way that its output marginal approaches a delta function at the data and its parameter marginal approaches the posterior distribution.
Inspired by Simulated Annealing, we present a new class of particle algorithms for ABC, based on a sequence of Metropolis kernels, associated with a decreasing sequence of tolerances w.r.t. the data.
Unlike other algorithms, our class of algorithms is {\em not} based on importance sampling. Hence, it does not suffer from a loss of effective sample size due to re-sampling.
We prove convergence under a condition on the speed at which the tolerance is decreased.
Furthermore, we present a scheme that adapts the tolerance and the jump distribution in parameter space according to some mean-fields of the ensemble, which preserves the statistical independence of the particles, in the limit of infinite sample size.
This adaptive scheme aims at converging as close as possible to the correct result with as few system updates as possible via minimizing the entropy production of the process.
The performance of this new class of algorithms is compared against two
other recent algorithms on two toy examples as well as on a real-world example from genetics.

\end{abstract}

\section{Introduction}
One way of implementing parameter inference in the Bayesian framework is to generate parameter samples from the {\em posterior distribution}
\begin{equation}\label{posterior}
f_{post}(\bt|\vc y)=\frac{f(\vc y|\bt)f(\bt)}{f(\vc y)}\,,
\end{equation}
where $f(\bt)$ denotes the {\em prior distribution} encoding our knowledge about the parameter vector $\bt$ before the experiment, $f(\vc y|\bt)$ is the {\em likelihood function}, that is, the probability density of outputs given the parameter vector $\bt$, evaluated at the measurement vector (data) $\vc y$, and $f(\vc y)$ is the corresponding prior density of the data.
Numerical methods such as the {\em Metropolis} algorithm \cite{metropolis_1953_MRT2} require many evaluations of the likelihood function to generate such a sample.
However, for
complex stochastic models, the likelihood function is often prohibitively
expensive to evaluate.
Therefore, in recent years, algorithms have been suggested that generate
samples from (\ref{posterior})
by {\em sampling model outputs} from the likelihood and comparing them with the data rather than evaluating the likelihood.

As far as we know, the origin of these algorithms is to be found in population genetics.
Tavar\'e et al. \cite{tavare_1997_ABC} replaced the output of a genetic model by a summary statistic and adopted a rejection technique to generate samples from the posterior.
Weiss et al. \cite{weiss_1998_ABC} extended this method sampling a vector of summary statistics and introducing a {\em tolerance} for its distance from the observed summary statistics. Thus, their algorithm generates samples from an {\em approximate} posterior.
Algorithms that generate samples from an approximate posterior via sampling outputs from the likelihood are nowadays called {\em Approximate Bayes Computations} (ABC).
Marjoram et al. \cite{marjoram_2003_ABC} used {\em Markov chains} to produce samples from an approximate posterior. Their algorithm combines a random walk in parameter space with drawing from the likelihood and an acceptance/rejection step that accounts for the prior and only accepts moves into an $\epsilon$ ball around the target $\vc y$. However, a small static tolerance leads to a high rejection rate.
Therefore, Toni et al. \cite{toni_2009_ABC} suggested using a decreasing sequence of tolerances and letting an ensemble of particles of constant size $N$ evolve towards an approximate posterior.
Their algorithm consists of an iteration of {\em importance sampling} steps, where each iteration consists of drawing a new ensemble from the old one with weights and subsequent re-sampling.  This re-sampling leads to a loss of effective sample size at each iteration step.
There are several adaptive versions of ensemble (or particle) ABC algorithms.
Beaumont et al. \cite{beaumont_2009_aABC} use the empirical variances of the ensemble to adapt the jump distribution in parameter space.
Del Moral et al. \cite{del2012adaptive} and Lenormand et al. \cite{Lenormand_2012_ABC} use the particles' distance from the target to adapt the tolerance.
Recent variants of the algorithm of del Moral et al appeared in \cite{lee_2012_ABC} and \cite{sedki_2013_ABC}.
All of the mentioned algorithms generate samples from the probability distribution proportional to $f(\bt)f(\vc x|\bt)\chi(\epsilon-\rho(\vc x,\vc y))$, where $\rho$ is some metric on the output space and $\chi$ denotes the Heaviside function whose value is unity if its argument is non-negative and 0 otherwise.
The effect of kernels different from the Heaviside function has been considered, e.g., in \cite{wilkinson_2007_ABC}.
For a recent review on ABC algorithms, the reader is referred to \cite{marin_2012_ABC}.

In this paper, we present a new class of (adaptive) ensemble algorithms that are of order $\mathcal O(N)$ and do not suffer from a loss of effective sample size.
The idea is to start with an ensemble of particles drawn from an arbitrary
distribution (e.g. the prior) in the product space of parameters and
outputs and apply a sequence of Markov kernels,
$(P_{\epsilon_k})$, each of which having
$$
Z^{-1}(\epsilon_k)f(\bt)f(\vc x|\bt)e^{-\rho(\vc x,\vc y)/\epsilon_k}
$$
as equilibrium distribution.
The key question is then how fast we should decrease $\epsilon_k$ in order to have a fast convergence and at the same time not to acquire an additional bias due to a too fast convergence. This problem is reminiscent of Simulated Annealing, which is one of our sources of inspiration.
We will give a convergence proof for a schedule that satisfies
$$
\epsilon_k\geq \const k^{-\alpha/n}\,,
$$
where $n$ is the dimension of the output space and $\alpha > 0$ is defined in (\ref{alpha}).
Furthermore, we will present an adaptive schedule that attempts convergence to the correct posterior while minimizing the required simulations from the likelihood.
Both the jump distribution in parameter space and the tolerance $\epsilon$ are adapted using mean fields of the ensemble.

The adaptation of $\epsilon$ we suggest is motivated from non-equilibrium thermodynamics, where this control parameter is naturally interpreted as a temperature.
We adapt $\epsilon$ according to the particles' distance to the target (energy) in such a way that the entropy production in the system, which is a measure for the waste of computation, is minimized.
A first order approximation of the entropy production is calculated using the so-called {\em endoreversibility assumption}, which states that the system undergoes only reversible changes, and which is approximately satisfied if either the mixing in parameter space is fast enough or annealing is slow enough. Under this assumption the only source of entropy production is the flow of energy (or rather heat) from the system to the environment, the latter being defined by the control parameter $\epsilon$ that is used for the transitions and can be interpreted as the temperature of a heat reservoir the system is in contact with.
In cases where the influence of the prior on the posterior is strong, we actively control this prior influence with a second control parameter, which allows us to extend the scope of the endoreversibility assumption.
Necessary and sufficient conditions for the minimization of entropy production, for endoreversible processes, have been derived in \cite{spirkl_1995_optFiniteTimeEndorevProc}.
For sufficiently slow processes, for which a linearity assumption holds, the condition is a {\em constant entropy production rate} \cite{salamon_1980_optHeatEngine}, which has been applied to Simulated Annealing, e.g., in \cite{ruppeiner_1991_EnsembleSA}.
In cases where the prior influence on the posterior is small, we go beyond the linearity assumption and suggest a scheme with non-constant entropy production rate.


The tolerance $\epsilon$ that can be achieved in reasonable time is limited by the dimension of the output space. This deficiency is inherent to all ABC algorithms simply because drawing an output from an $\epsilon$-ball around $\vc y$ scales like $\epsilon^n$.
Methods to reduce this bias are investigated elsewhere (see, e.g., \cite{Fearnhead_2012_ABC}, \cite{leuenberger_2010_ABC}).

The paper is organized as follows:
In Subsect. \ref{SectBasic}, we explain the main idea behind our class of algorithms.
In Subsect. \ref{Sectexplicit}, the explicit scheme together with a convergence proof is given.
The adaptive scheme is developed in Subsect. \ref{SectCTS}.
Sect. \ref{Toys} contains an application to two toy models, for which the posterior is available analytically, as well as a comparison with two recent adaptive ABC algorithms \cite{del2012adaptive}, \cite{Lenormand_2012_ABC}.
Sect. \ref{CaseStudy} contains an application in genetics.
Conclusions are drawn in Sect. \ref{SectConclusions}.

\section{A new class of ABC algorithms}\label{Algorithm}
\subsection{Basic idea}\label{SectBasic}
Our aim is to sample from the posterior distribution (\ref{posterior}),
without evaluating the likelihood function. The basic idea behind ABC
is to rewrite (\ref{posterior}) as the marginalization
\bb{marginalization}
f_{post}(\bt|\vc y)\propto \int f(\vc x|\bt)f(\bt)\delta(\vc x-\vc y)d\vc x\,
\ee
and sample from the joint density $ f(\vc x|\bt)f(\bt)\delta(\vc x-\vc
y)$ in the $(\bt,\vc x)$-space, $\Theta\times X$, which means to sample a parameter vector from the prior and an associated output from the likelihood and accept the particle iff the drawn output happens to coincide with the data.
If the output space has a high cardinality or is continuous, sampling from
$f(\vc x|\bt)f(\bt)\delta(\vc x-\vc y)$ becomes inefficient or impossible,
respectively. In these cases, we approximate it by the following
family of distributions
\bb{pi1}
\pi_{\epsilon}(\bt,\vc x)=
\frac{1}{Z(\epsilon)}
f(\vc x|\bt)f(\bt)e^{-\rho(\vc x,\vc y)/\epsilon}\,,
\ee
where $\rho(\vc x,\vc y)$ measures how close $\vc x$ is to the observation $\vc y$.
For simplicity, we set $X=\mathbb R^n$ and
\begin{equation}\label{alpha}
  \rho(\vc x,\vc y)
  =
  \frac{1}{\alpha}
  \sum_{i=1}^n|x_i-y_i|^\alpha\,,
\end{equation}
for some $\alpha > 0$, but our results could easily be extended to more general  manifolds equipped with distance measures obeying suitable regularity conditions.
This might become necessary if {\em summary statistics} are used to map the output space to some smaller-dimensional manifold (see, e.g., \cite{Fearnhead_2012_ABC}, \cite{tavare_1997_ABC} and \cite{weiss_1998_ABC}).

Under the assumption that $f(\vc x|\bt)$ is uniformly bounded and, as a
function of $\vc x$, continuous at $\vc y$, $\pi_\epsilon$ converges
weakly to $f(\vc x|\bt)f(\bt)\delta(\vc x-\vc y)d \bt d \vc x$,
for $\epsilon\searrow 0$.
Our idea is to choose a
family of Markov transition kernels
$(P_\epsilon)$ on the space $\Theta\times X$, which
have $\pi_{\epsilon}$ as stationary distribution
and apply them recursively on members of a sample drawn from an arbitrary initial distribution, for a decreasing sequence of $\epsilon$'s.
If $\epsilon$ is decreased sufficiently slowly, we expect to end up with an approximate sample from the posterior distribution.
This is analogous
to the Simulated Annealing algorithm, although in Simulated Annealing
the limiting distribution is usually concentrated on a finite set.
Still, we will strongly
rely on ideas developed in the context of Simulated Annealing.
The transition kernels $(P_\epsilon)$ that we will use in Subsects. \ref{Sectexplicit} and \ref{nlSABC} are defined by the transition densities
\begin{equation}\label{transition}
q_\epsilon((\bt',\vc x'),(\bt,\vc x))=k(\bt',\bt)f(\vc x|\bt)
\min
\left(1,
	\frac{f(\bt)e^{-\rho(\vc x,\vc y)/\epsilon}}
		{f(\bt')e^{-\rho(\vc x',\vc y)/\epsilon}}
\right)\, ,
\end{equation}
combined with a multiple of a Dirac delta distribution at $(\bt',\vc x')$ such that
$P_\epsilon((\bt',\vc x'),\Theta\times X)=1$. Here, $k$ is a symmetric
transition density on $\Theta$. It is straightforward to
check that $\pi_{\epsilon}$ is the equilibrium distribution for $P_\epsilon$.

The main question now is how fast $\epsilon$ should be decreased.
Obviously, an arbitrarily slow decrease of $\epsilon$ allows to stay arbitrarily close to equilibrium at all times after, possibly, an initial burn-in period, which guarantees convergence. However, this is clearly inefficient.
On the other hand, a too fast decrease may result in slow convergence (because
the acceptance probability decreases for decreasing $\epsilon$) or
convergence to a biased result.
A bias can occur, e.g., if the prior within the last factor in eq. (\ref{transition}) decides too seldom whether a proposal point in $\Theta\times X$ is accepted or not.
In the extreme case of a constant $\epsilon=0$, the acceptance term in (\ref{transition}) becomes $\chi(\rho(\vc x',\vc y)-\rho(\vc x,\vc y))$, thus, $(\bt, \vc x)$ is accepted
iff $\rho(x,y) \leq \rho(x',y)$. Hence in this case, the prior has no
influence, which clearly leads to convergence to a biased result.
For this reason, in Subsect. \ref{SectlinSABC}, we will introduce a second control parameter to control the influence of the prior and replace (\ref{transition}) by (\ref{newtransition}).

In the next subsection we will present an explicit schedule $(\epsilon_k)$ that ensures convergence to an unbiased result.
A potentially better performance can be achieved when the state of the system is used to adapt the tolerance $\epsilon$ and the jump distribution $k$. This idea will be developed in Subsect. \ref{SectCTS}.

\subsection{An explicit scheme with convergence proof}\label{Sectexplicit}

In this subsection, we use a time discrete description.
That is, we start with a sample from an arbitrary distribution $\mu_0$ and then recursively make transitions of the whole sample with the kernel $P_{\epsilon_k}$, for an explicitly given decreasing sequence $\epsilon_k\searrow 0$.
In this way, we generate
samples distributed according to
\begin{equation}\label{discTransition}
\mu_{k+1} =\mu_kP_{\epsilon_{k+1}} =
\int P_{\epsilon_{k+1}}(\bt,\vc x; .) d\mu_k(\bt,\vc x).
\end{equation}
We expect that for a suitable choice
of $(\epsilon_k)$, $\mu_k$ will converge weakly to
$f(\vc x|\bt)f(\bt)\delta(\vc x-\vc y) d \bt d \vc x$, and thus
in particular the marginal will converge weakly to the posterior distribution (\ref{posterior}).

In order to ease the notation we set $\vc z=(\bt^T,\vc x^T)^T$ and write,
for the joint prior,
$$
f(\vc z):=f(\vc x|\bt)f(\bt)\,.
$$
Furthermore, w.l.o.g. we will assume $\vc y=\vc 0$ and replace $\rho(\vc x,\vc y)$ by
$\rho(\vc x)$.
For our main result, we make the following assumptions about the parameter
space $\Theta$ and the functions $k(\bt',\bt)$, $f(\bt)$ and
$f(\vc x|\bt)$ thereon:
\begin{enumerate}
\item[(A1)]
$\exists c_1>1$ such that $c_1^{-1}\leq f(\bt)/f(\bt')\leq c_1$, for all $\bt,\bt' \in \Theta$.
\item [(A2)]
$\exists c_2>0$ such that $k(\bt',\bt)\geq c_2 f(\bt)$, for all $\bt, \bt'
\in \Theta$.
\item[(A3)] $f(\vc x|\bt)$ is continuously differentiable w.r.t. $\vc x$
for all $\bt$,
and the function and all partial derivatives are bounded uniformly in
$\vc x$ and $\bt$.
\end{enumerate}
These conditions essentially restrict the parameter space to be compact.
We will in fact prove stronger
than weak-convergence results, namely convergence in total
variation of the distributions of $(\bt, \epsilon_k^{-1/\alpha}\vc x)$, with $\alpha>0$ as defined in (\ref{alpha}).
The densities of these scaled distributions are
$$
\hat\mu_k(\bt,\vc x)
:=
\epsilon_k^{n/\alpha}\mu_k(\bt,\epsilon_k^{1/\alpha}\vc x)$$
and
$$
\hat\pi_{\epsilon}(\bt,\vc x)
:=
\epsilon^{n/\alpha}\pi_\epsilon(\bt,\epsilon^{1/\alpha}\vc x)
= \frac{1}{C(\epsilon^{1/\alpha})} f(\epsilon^{1/\alpha} \vc x|\bt)f(\bt)\exp(-\rho(\vc x))\,,
$$
where
$$C(\epsilon^{1/\alpha}) = \int f(\epsilon^{1/\alpha} \vc x|\bt)f(\bt) \exp(-\rho(\vc x) )
d \vc z\,,$$
and the transition densities for the scaled variables are
$$
\hat q_{\epsilon^{1/\alpha}}(\vc z,\vc z')=\epsilon^{n/\alpha}
q_\epsilon((\bt,\epsilon^{1/\alpha}\vc x),(\bt',\epsilon^{1/\alpha}\vc x'))\,.
$$

\begin{theorem}\label{Thm}
If the assumptions (A1) -- (A3) above are satisfied and if
\bb{coolingexplicit}
\epsilon_k \geq \const k^{-\alpha/n}\,,
\ee
for an arbitrary constant (where $n$ denotes the dimension of $X$ and $\alpha$ is defined by (\ref{alpha})),
then, for any absolutely continuous initial distribution $\hat \mu_0$
the distribution $\hat\mu_k$ converges in total variation to
$\hat\pi_0(\vc z)\propto f_{post}(\bt|\vc y)\exp(-\rho(\vc x))$, for
$k\rightarrow\infty$.
\end{theorem}

{\bf Proof:} We will apply corollary (2.34)
in \cite{foellmer_1988_RandomFields}. We start by introducing
some notation. Let
$$
\hat\pi_k=\hat\pi_{\epsilon_k}\,,\quad \hat P_k=\hat P_{\epsilon_k}\,,\quad \hat P_{s:t}=\hat P_s\hat P_{s+1}\dots\hat P_t\,,
$$
where $\hat P_\epsilon$ is defined by the transition density $\hat q_\epsilon$.

By assumption (A3) and dominated convergence,
$$
\hat{\pi}_k(\bt, \vc x) \rightarrow \hat{\pi}_0(\bt, \vc x) =
\frac{f(\vc 0|\bt) f(\bt) \exp(-\rho(\vc x))}
{\int f(\vc 0|\bt)f(\bt) d\bt \int \exp(-\rho(\vc x)) d\vc x}
$$
pointwise and thus by Scheff\'e's theorem also in $L^1$-norm,
that is in total variation. In order to deduce
$$ ||\hat{\mu}_0 \hat{P}_{0:t} - \hat{\pi}_0||_{TV} \rightarrow 0,$$
we have to verify conditions (2.31) and (2.33) in
\cite{foellmer_1988_RandomFields}. These conditions are
\bb{cond1}
\prod_k c(\hat{P}_k) = 0\,,
\ee
where
$$
c(\hat P_k)=\underset{\vc z,\vc z'}{\operatorname{sup}}
||\hat P_k(\vc z,.)-\hat P_k(\vc z',.)||_{TV}\,,
$$
and
\bb{cond2}
\sum_k ||\hat{\pi}_{k+1} - \hat{\pi}_k||_{TV} < \infty.
\ee

Replacing $\epsilon^{1/\alpha}$ by $\epsilon$, we may set, without loss of generality, $\alpha=1$.
To get an upper bound for $c(\hat{P}_\epsilon)$ we use
$$c(\hat P_\epsilon) =
\underset{\vc z',\vc z''}{\operatorname{sup}}
\left(1- \int \operatorname{min}(\hat q_\epsilon(\vc z',\vc z),
\hat q_\epsilon(\vc z'',\vc z))d\vc z \right).$$
By (A1) and (A2), for any $\vc z'$,
$$
\hat q_\epsilon(\vc z',\vc z)
\geq
\epsilon^{n}
\frac{c_2}{c_1} f(\bt) f(\epsilon \vc x | \bt)\exp(-\rho(\vc x))\,.
$$
Hence we obtain
$$
\int \operatorname{min}(\hat q_\epsilon(\vc z',\vc z),
\hat q_\epsilon(\vc z'',\vc z))d\vc z \geq
\epsilon^{n}\frac{c_2}{c_1} C(\epsilon).
$$
Because $C(\epsilon) \rightarrow C(0) > 0$ as $\epsilon \rightarrow 0$,
it follows that, for $\epsilon$ sufficiently small $\epsilon$,
\bb{ineq1}
c(\hat P_\epsilon) \leq 1-\frac{c_2}{c_1} \frac{C(0)}{2} \epsilon^{n}\, ,
\ee
and (\ref{cond1}) holds for the choice (\ref{coolingexplicit}).

In order to show (\ref{cond2}), we start with
$$
|\hat\pi_\epsilon(\vc z) - \hat\pi_{\epsilon'}(\vc z)|
\leq \frac{|f(\epsilon \vc x|\bt) - f({\epsilon'} \vc x|\bt)|
f(\bt)\exp(-\rho(\vc x))}{C(\epsilon)} + \hat\pi_{\epsilon'}(\vc z)
\frac{|C(\epsilon') - C(\epsilon)|}{C(\epsilon)}\,.
$$
By (A3) and the intermediate value theorem, we obtain that
$$
|f(\epsilon \vc x|\bt) - f({\epsilon'} \vc x|\bt)|
\leq
\const  ||\vc x||_1|\epsilon - {\epsilon'}|
$$
and, moreover, that $C(\epsilon)$ is differentiable with
$$|C'(\epsilon)| \leq \const \int ||\vc x||_1 \exp(-\rho(\vc x)) d\vc x\,,$$
where $\const$ is the bound for the partial derivatives of $f(.|\bt)$.
Hence we find that
$$||\hat\pi_{\epsilon} - \hat\pi_{\epsilon'}||_{TV} \leq
\frac{\const}{C(\epsilon)} \int ||\vc x||_1 \exp(-\rho(\vc x)) d\vc x
\, |\epsilon - \epsilon'|\,.$$
Therefore (\ref{cond2}) holds for any sequence $(\epsilon_k)$
which converges monotonically to zero.

\hfill $\Box$

{\bf Remark:}
Convergence of inhomogeneous Markov chains has been proved in much more general settings than in \cite{foellmer_1988_RandomFields}, see e.g. \cite{douc_2004_inhomoMC}, or Proposition A.1 in \cite{Beskos_2012_SMC}.
Using these techniques, it should be possible to relax the assumptions (A1)--(A2).

\subsection{An adaptive scheme}\label{SectCTS}

\subsubsection{Heuristics}

As stated in Section \ref{SectBasic}, we construct an ensemble of particles
which evolve according to a family of Markov transition
kernels $(P_\epsilon)$ with a control parameter
$\epsilon=\epsilon^e(t)$ that decreases to zero (the reason for the
notation $\epsilon^e(t)$ will become clear later). In contrast to the
algorithms in \cite{del2012adaptive} and \cite{Lenormand_2012_ABC},
we do not use importance sampling
to force the distribution of the ensemble to agree with the target
distribution (\ref{pi1}) at certain time points.
This has the advantage that the effective sample size of the ensemble does
not decrease over time, but the disadvantage that we loose control over the
transient distribution of the stochastic process defined by the algorithm.
However, as  Theorem 2.1 suggests, this distribution remains close to an
equilibrium (\ref{pi1}) at all times, if either the value of the
control parameter $\epsilon^e(t)$ is lowered
sufficiently slowly or if mixing in parameter space is sufficiently fast.
In this section, we shall design an algorithm that adapts $\epsilon^e(t)$
based on the average distance of the particles from the target
$\vc y = \vc 0$ in such a way
that the computational effort, that is the number of draws from the
likelihood, is minimized.  There is therefore a mean field interaction
between particles.

The design of the algorithm will rely on the {\em assumption} that the
distribution of the Markov chain is at all times $t$ close to an
equilibrium distribution $\pi_{\epsilon(t)}$, but with a parameter $\epsilon(t)$
which is somewhat higher than the value $\epsilon^e(t)$
used for the transition. How quickly we let $\epsilon^e(t)$ go to zero
as the algorithm proceeds is our decision, and it determines together with
the jump distribution $k(\bt,\bt')$ in parameter space the function
$\epsilon(t)$. We have no analytical expression for $\epsilon(t)$,
but it is in a one-to-one correspondence with the expected distance, $U(t)$,
from the target, that we can estimate.

Since the intuition behind our adaptive algorithm stems from
non-equilibrium thermodynamics, it might be helpful to imagine a gas,
which is in contact with a heat bath whose {\em temperature} $\epsilon^e(t)$ can
be controlled. The value of $\epsilon(t)$ is then the temperature of the
gas at time $t$ which is measured continuously and influences how quickly
$\epsilon^e(t)$ is lowered. The superscript $e$ stands for {\em "environment"} or
{\em "equilibrium"}, because it defines the equilibrium state the system would
relax to if cooling suddenly stopped, that is if $\epsilon^e(t)$ would be
kept constant after some time $t_0$.  However, if the temperature
$\epsilon^e(t)$ is continuously lowered, then the gas will at any time $t$
be warmer than the environment. In the physics community, a system which is
always described by an equilibrium distribution even if it is externally
driven, i.e., never at equilibrium with its environment, is called
{\em endoreversible} (see \cite{rubin_1979_endoreversibility}).
The system is then described by the Gibbs state $\pi_{\epsilon(t)}$, and the distance of a particle to the target is interpreted as the particle's {\em energy}.

The question is then how $\epsilon^e(t)$ should be controlled, depending on
the distribution of the system given by $\epsilon(t)$ or $U(t)$, so as to
waste as little computation as possible.  In physics' terms, the cooling of
the system by lowering the temperature of the environment creates a {\em flow of
entropy} from the system to the environment.
It can  be split into two parts.  One part is the (path-independent) reduction of the
system's entropy. This is the well invested part of the computing effort,
as it measures the information difference between prior and posterior.  The
other part is the {\em entropy production}, which is a measure for the wasted computing effort.
We argue therefore that
we have to choose the cooling or annealing schedule $\epsilon^e(t)$
such that this entropy production is minimized. Using variational calculus
\cite{spirkl_1995_optFiniteTimeEndorevProc}, this approach leaves
us with a family of  annealing schedules, parameterized by a tuning
parameter $v$, which governs the annealing speed and expresses the optimal $\epsilon^e(t)$ in function of the expected distance $U(t)$ from the target.

In mathematical terms, entropy production equals the {\em Shannon entropy} of the
probability  distribution of the process, on the space of paths, relative to the
time-inverted stochastic process \cite{seifert_2005_EntropyProductionMasterEq}.
It can be seen as a measure for the information loss due to rejections: With a fast cooling
schedule, a typical path is likely to encounter many more rejections than a
typical path under the time-reversed (heating) schedule. Thus, the
probability distribution of the stochastic process on the space of all
paths is in this case much more concentrated than the distribution of the
reverse process, which leads to a large relative entropy.

The assumption of endoreversibility is crucial for our algorithm.
If it is violated, there will be additional production
of entropy due to irreversible processes.  This entropy
production is beyond control if only the energy of the system is measured
and is thus to be avoided.  Such additional entropy can even remain in the
system indefinitely and lead to a biased convergence.
Whether the assumption of endoreversibility is justified or not will depend
on the values of the two tuning parameters of the algorithm: The covariance
$K$ of the jump distribution $k(\bt,\bt')$ in parameter space and the
tuning parameter $v$ that arises from variational calculus.
A small
value of $K$ leads to slow mixing in parameter space, and if it is too
slow compared to the decay rate of $\epsilon(t)$, the endoreversibility
assumption might be violated. We derive our cooling schedule
under the assumption that $K$ is constant, but in practice it is
usually advantageous to adapt $K$ to the current distribution of
the chain. We will discuss the adaptive choice of $K$ at the end of
the next subsection.

Similarly, a too large $v$ can lead to a too fast cooling, compared to the mixing in parameter space, which bears the
risk of violating  the endoreversibility
assumption. On the other hand, a too small value of $v$ leads to a
large amount of reversible computations which is not accounted for by
the entropy production. Running the algorithm at equilibrium, i.e. setting
$\epsilon^e(t)=\epsilon(t)$, does neither lead to a flow of entropy nor
does it generate any entropy and would thus be considered optimal
by our criterion. Because $v$ has the dimension  of an inverse time,
measured in units of $N$ computer updates of single particles, its optimal
value is expected to depend not too much on details of the model.

The problem of choosing a good value of $v$ is pronounced if the prior,
$f(\bt)$, carries relevant
information.  Since $\epsilon^e(t)$ is by construction of the algorithm
smaller than the value $\epsilon(t)$ implied by the expected
distance from the
target, the $\rho(\vc x,\vc y)$-dependent term in (\ref{transition}) tends
to decide more often than the prior-dependent term whether or not to accept
a move. This may lead to an under-representation of the prior in the final
solution.  Thus, if the prior is important, we suggest introducing a second
pair of ``temperatures'' $\epsilon^e_2(t)$ and $\epsilon_2(t)$, in order to
control the relative information
contributed by the prior.  Obviously, in this case, tuning will become much
more sophisticated and we only derive an optimal schedule for relatively
slow annealing, in which case the relation between forces and fluxes (to be
defined below) is approximately linear.  The more information the prior
contains, however, the less advantageous a sequential scheme as ours
appears compared to a brute-force acceptance/rejection algorithm.
Therefore, we devote the next subsection to the simpler, yet practically
relevant, case of negligible prior information.

\subsubsection{The case of negligible prior information}\label{nlSABC}

In this subsection we consider the special case where the prior $f(\bt)$
doesn't play much of a role. This is the case if $f(\bt)\approx \const$, in
the area where the likelihood function, evaluated at the data $\vc y$, is
not negligible.

Our {\em system} is an inhomogeneous continuous time
Markov process $(\vc Z_t)$ on the product space of parameters and model
outputs. Its transitions occur at the random
times of a Poisson process with rate 1, according to the transition kernel
(\ref{transition})
with time dependent parameter $\epsilon^e(t)$. This means that the density
$\mu(\vc z,t)$ of the system at time $t$ satisfies
\begin{equation}
\label{time-evol-mu}
\frac{\partial \mu(\vc z,t)}{\partial t}
= \int \mu(\vc z',t) q_{\epsilon^e(t)}(\vc z', \vc z) d\vc z' - \mu(\vc z,t)
\int q_{\epsilon^e(t)}(\vc z, \vc z') d\vc z'\, ,
\end{equation}
and for functions $h$ on the product space we have
\begin{equation}
\label{time-evol-h}
\frac{d E(h(\vc Z_t))}{d t}
= \int (h(\vc z) - h(\vc z'))\mu(\vc z',t) q_{\epsilon^e(t)}(\vc z', \vc z)
d\vc z d\vc z'\,.
\end{equation}
The parameter $\epsilon^e$ which controls the cooling
of the system is adaptive in the sense
that $\epsilon^e(t)$ depends on the distribution $\mu(\vc z,t)$.
In our algorithm, we will represent the system by a sufficiently large
{\em ensemble}, $E$, of particles, $\{\vc z_i=(\bt_i,\vc x_i)\}_{i=1}^N$
which evolve in time. Each system update consists in choosing a
random member of the ensemble and updating it according to the transition
kernel (\ref{transition}). The parameter $\epsilon^e(t)$ is then
based on the current empirical distribution of the ensemble at time $t$.

As discussed above we will assume the process to satisfy the
endoreversibility assumption
\begin{equation}\label{muapprox}
  \mu(\vc z,t)\approx \pi_{\epsilon(t)}(\vc z)\,,
\end{equation}
where $\pi_\epsilon(\vc z)$ was defined in (\ref{pi1}).
As we have discussed the legitimacy of this assumption in the previous
subsection, we take it for granted here. The system's
temperature $\epsilon(t)$ is in one-to-one correspondence with the
system's energy which is the system's expected distance to the target.
It will be measured by the average distance of the particles from the
target.

We derive now our algorithm for the choice of the cooling schedule
$\epsilon^e(t)$ in a sequence of steps. In the first step we modify
distance by a monotone transformation to get approximate equality of
energy and temperature. We define
\begin{equation}\label{newmetric}
u(\vc x) = G(\rho(\vc x)), \quad G(\rho)=\int_{\rho(\vc x) \leq \rho}
f(\bt,\vc x)d\bt d\vc x\,,
\end{equation}
and we replace $\rho(\vc x)$ by $u(\vc x)$ in the definitions of $\pi_{\epsilon}$
and $q_{\epsilon}$.
Because $G$ is the cumulative distribution function of $\rho(\vc x)$ under
the prior $f(\vc x,\bt)$, we obtain, for the mean energy
under $\pi_{\epsilon}$,
\begin{equation}\label{Up}
    U(\epsilon):=\int u(\vc x)\pi_{\epsilon}(\bt, \vc x)d\bt d\vc x\,,
\end{equation}
the expression
\begin{equation}
  U(\epsilon) = \frac{\int_0^\infty G(\rho)e^{-G(\rho)/\epsilon}
  G'(\rho)d\rho}{\int_0^\infty e^{-G(\rho)/\epsilon} G'(\rho)d\rho}
   =  \frac{\int_0^1 u e^{-u/\epsilon} du}{\int_0^1 e^{-u/\epsilon} du}
   = \epsilon \frac{1-e^{-1/\epsilon}(1+ 1/\epsilon)}{1-e^{-1/\epsilon}}\,.
\end{equation}
As $\epsilon$ goes to zero, the fraction on the right is
$1+o(\epsilon^k)$ for any $k>0$. By the endoreversibility assumption we
therefore have
\begin{equation}\label{U}
  U(t) := \int u(\vc x) \mu(\vc z,t) \approx U(\epsilon(t)) \approx
\epsilon(t)\,.
\end{equation}

Our main result of this section, equation (\ref{cooling}) below,
expresses the optimal cooling schedule $\epsilon^e(t)$ as a function
of $U(t)$ and a tuning parameter $v$ of the algorithm. In order to
estimate $U(t)$ we need first an approximation of the distribution
function $G$ which we construct at the beginning of the
algorithm, based on the prior sample, $P$, that is drawn to get the initial
ensemble $E$. If the sample size of $P$ is not large enough or if we want
to run the algorithm for a very long time, we might want to use a smooth
approximation  of the empirical distribution of the values
$\rho(\vc x_i)$, which, for small $\rho$, and
for $\alpha=2$ in (\ref{alpha}), behaves as
\begin{equation}
  G(\rho) \approx \const \rho^{n/2}\,.
\end{equation}

In the second step, we approximate $\dot U(t) = \frac{d}{dt}U(t)$, the
so-called flux, as a
function of $\epsilon(t)$ and $\epsilon^e(t)$. For this, we cannot
use directly $U(t) \approx U(\epsilon(t))$ because then the dependence
on $\epsilon^e(t)$ would be lost. Combining the endoreversibility
assumption with the time evolution (\ref{time-evol-h}) gives
\begin{eqnarray}
  \dot U(t) &=& \int (u(\vc x)-u(\vc x'))k(\bt',\bt)f(\vc x|\bt) \min\left(
    1,\frac{f(\bt)\exp(-u(\vc x)/\epsilon^e(t))}{f(\bt')
  \exp(-u(\vc x')/\epsilon^e(t))}\right) \mu(\vc z',t) d\vc z d\vc z'\nonumber\\
  &\approx& Z^{-1}(\epsilon(t))
  \int (u(\vc x)-u(\vc x'))k(\bt',\bt)f(\vc x|\bt)f(\vc x'|\bt') \nonumber\\
  & & \times \min\left(f(\bt'),f(\bt) \frac{\exp(-u(\vc x)/\epsilon^e(t))}
  {\exp(-u(\vc x')/\epsilon^e(t))}\right) \exp(-u(\vc x')/\epsilon(t))
   d\vc z d\vc z'\,. \label{Udot}
\end{eqnarray}
Since $\epsilon$ (and thus also $\epsilon^e$) will be much smaller than $1$
during most of the process, we use a Taylor expansion of (\ref{Udot})
to quadratic order in $\epsilon$ and $\epsilon^e$. Under the assumption
that the influence of the prior is negligible, we obtain
\begin{equation}\label{Udot10}
  \dot U(t) \approx \dot U(\epsilon,\epsilon^e)
  \approx
  -\gamma(\epsilon^2-(\epsilon^e)^2)\,,
\end{equation}
with
\begin{equation}\label{Udot20}
  \gamma
  =
  (f(\vc y))^{-2}
  \int k(\bt',\bt)f(\vc y,\bt)f(\vc y,\bt')d\bt d\bt'
 \,.
\end{equation}
For later use, we note that from $U(t) \approx \epsilon(t)$ we obtain
\begin{equation}
  \label{eps-e-1}
  \epsilon^e(t) \approx  \sqrt{U(t)^2 + \dot U(t)/\gamma}\,.
\end{equation}

In the third step we approximate the derivative of the irreversible
process entropy or entropy production. To simplify the notation, let us begin with the
version in discrete time where we have an initial distribution
$\mu_0$ and a sequence of transition kernels $P_i$ corresponding
to a sequence $\epsilon^e_i$ of control parameters. The probability
of a path $\Gamma_n = (\vc z_0,\vc z_1,\dots,\vc z_{n-1})$ is then
\begin{equation}
p(\Gamma_n)
=
\mu_0(\vc z_0)P_0(\vc z_0,\vc z_1)\dots P_{n-2}(\vc z_{n-2},\vc z_{n-1})\,,
\end{equation}
whereas the probability of the same path with respect to
the time-reverse schedule is
\begin{equation}
p^R(\Gamma_n) =
\mu_{n-1}(\vc z_{n-1})P_{n-2}(\vc z_{n-1},\vc z_{n-2})\dots P_{0}(\vc z_1,\vc z_0)\,,
\end{equation}
where
$$\mu_{n-1}(\vc z_{n-1}) = \int p(\Gamma_n) d\vc z_0 \cdots d\vc z_{n-2}$$
is the distribution of the final state. The irreversible process entropy is
then defined as the relative entropy of $p^R$ with respect to $p$,
see \cite{seifert_2005_EntropyProductionMasterEq},
\begin{equation}
S_{irr}(n) = \int p(\Gamma_n) \ln
	\frac{p(\Gamma_n)}{p^R(\Gamma_n)} d\Gamma_n\,.
\end{equation}
From this it follows easily that
\begin{multline}
S_{irr}(n+1) = S_{irr}(n) \\
+ \int \log \left(
	\frac{\mu_{n-1}(\vc z_{n-1}) P_{n-1}(\vc z_{n-1},\vc z_{n})}
        {\mu_{n}(\vc z_{n}) P_{n-1}(\vc z_{n},\vc z_{n-1})}\right)
        \mu_{n-1}(\vc z_{n-1}) P_{n-1}(\vc z_{n-1},\vc z_{n}) d\vc z_{n-1}
        d \vc z_n \,.
\end{multline}
Passing to a continuous time limit, we therefore obtain from
(\ref{time-evol-mu})
\begin{multline}
\dot S_{irr}(t) = \int \log \left(
     \frac{q_{\epsilon^e(t)}(\vc z,\vc z')}{q_{\epsilon^e(t)}(\vc z',\vc
       z)}\right) \mu(\vc z,t) q_{\epsilon^e(t)}(\vc z,\vc z')d \vc z d\vc z'
    - \frac{d}{dt} \int \log(\mu(\vc z,t)) \mu(\vc z,t) d\vc z\\
    = \int \log \left(
     \frac{q_{\epsilon^e(t)}(\vc z,\vc z')}{q_{\epsilon^e(t)}(\vc z',\vc
       z)}\right) \mu(\vc z,t) q_{\epsilon^e(t)}(\vc z,\vc z')d \vc z d\vc z'
    - \int \log(\mu(\vc z,t)) \frac{\partial \mu(\vc z,t)}{\partial t}d\vc z\\
    =  \int \log \left(
     \frac{q_{\epsilon^e(t)}(\vc z,\vc z')\mu(\vc z,t)}
     {q_{\epsilon^e(t)}(\vc z',\vc z)\mu(\vc z',t)}\right)
    \mu(\vc z,t) q_{\epsilon^e(t)}(\vc z,\vc z')d \vc z d\vc z'\,.
\end{multline}
Using the endoreversibility assumption and the expression (\ref{transition})
for $q_\epsilon$, we arrive at
\begin{multline}
  \dot S_{irr}(t)\label{EntProdRate}
  =
  \int
     \mu(\vc z,t)q_t(\vc z,\vc z')
  (u(\vc z')-u(\vc z))
  \left(
    \frac{1}{\epsilon(t)} - \frac{1}{\epsilon^e(t)}
  \right)
  d\vc z d\vc z'\\
  =
  \left(
    \frac{1}{\epsilon(t)} - \frac{1}{\epsilon^e(t)}
  \right)
  \frac{d}{dt}
  \int u(\vc z)\mu(\vc z,t) d\vc z
  =
  F(t)\dot U(t)\,,
\end{multline}
where
\begin{equation}\label{1DForce}
  F(t) = \epsilon(t)^{-1}-\epsilon^e(t)^{-1}\,
\end{equation}
is the thermodynamic force, the difference between the inverse temperatures
of the system and the environment. Because of (\ref{U}) and (\ref{eps-e-1})
$F(t)$ is a function of $U(t)$ and $\dot U(t)$,

In the fourth step, we determine the necessary and sufficient criterion for minimal
entropy production, for fixed initial and final values of the energy:
$$\int_{0}^{t_f} F(U(t),\dot U(t))\dot U(t)dt=\min!, \quad
U(0)=U_0, \ \ U(t_f)=U_f.$$
Using standard methods of variational calculus, see
\cite{spirkl_1995_optFiniteTimeEndorevProc}, one obtains the differential
equation
\begin{equation}\label{preschedule}
 \dot U
 \frac{\partial F}{\partial \dot U}
 \dot U
 = \const =v\,.
\end{equation}
From (\ref{eps-e-1}) it follows that
$$\frac{\partial F}{\partial \dot U} =
- \frac{\partial \epsilon^e(t)^{-1}}{\partial \dot U}
\approx \frac{\gamma^{1/2}}{2(\gamma U(t)^2
  + \dot U(t))^{3/2}} \approx \frac{1}{2 \gamma \epsilon^e(t)^3}.$$
If we combine this result with (\ref{Udot10})
we find the optimal cooling schedule, for small $U$, to be approximated
by the unique solution of the quartic equation
\begin{equation}\label{cooling}
  \frac
  {(U(t)^2 - \epsilon^e(t)^2)^2}
  {2 \epsilon^e(t)^3}
  =
  \frac{v}{\gamma}\
\end{equation}
in the interval $(0,U(t))$. It can be computed efficiently with the Newton
algorithm. The leading term of the solution $\epsilon^e(U)$, for small $U$, is
\begin{equation}
  \epsilon^e(U)=\left({\frac{\gamma}{2v}}\right)^{1/3}
  U^{4/3}+\mathcal O(U^2)\,.
\end{equation}
This means that the cooling is slowing down when $U(t)$ gets small. One can
derive from this also an explicit cooling schedule,
\begin{equation}\label{fastcooling}
  \epsilon^e(t)\sim t^{-4/3}\,,
\end{equation}
but this will not be used in our algorithm. It shows however that the
cooling schedule which follows from Theorem \ref{Thm} is different from the adaptive
schedule here.

For convenience, the algorithm derived in this subsection is given as a
pseudo-code in Table \ref{AlgI}.

\begin{table}
\begin{framed}
{\bf Input:}
\begin{enumerate}
   \item Algorithms to sample from the prior and the likelihood.
   \item Ensemble size $N$ and initial value $\epsilon_{init}$.
   \item Covariance $K$ of the jump distribution
         $k(\bt,\bt') = \mathcal{N}(\bt, K)$.
   \item Tuning parameter $v$. The default value is $v=0.3$.
\end{enumerate}

{\bf Initialization:}
\begin{enumerate}
  \item
  Repeat, until the ensemble $E$ constructed in (d) contains $N$ particles:
  \begin{enumerate}
    \item
    Sample a parameter vector, $\bt$, from the prior.
    \item
    Sample an output, $\vc x$, from the likelihood $f(\vc x|\bt)$.
    \item
    Store the particle $(\bt, \rho(\vc x,\vc y))$ in the ensemble $P$.
    \item
    With probability $\exp[-\rho(\vc x,\vc y)/\epsilon_{init}]$ store the particle $(\bt,\rho(\vc x,\vc y))$ also in ensemble $E$.
  \end{enumerate}
  \item
  Estimate the distribution function $G=G(\rho)$ defined in
  (\ref{newmetric}) by smoothing the empirical distribution of $\rho(\vc
  x,\vc y)$ in the ensemble $P$, and re-calculate all the distances in
  ensemble $E$ as $u=G(\rho(\vc x, \vc y))$.
   \item
   Initialize $U$ as the average of the redefined distance $u$ in ensemble $E$.
    \item
  Estimate $\gamma$ defined in (\ref{Udot20}) using the prior ensemble $P$.
  \item
  Initialize $\epsilon^e$ solving the quartic equation (\ref{cooling}).
  \item
  Initialize $K$ according to (\ref{beta2}).
\end{enumerate}

{\bf Iteration:}
\begin{enumerate}
  \item
  Select a random particle, $(\bt,u)$, from the ensemble $E$.
  \item
  Sample a proposal parameter vector, $\bt^*$, from $k(\bt,\bt^*)$.
  \item
  Sample a proposal output, $\vc x^*$, from the likelihood $f(\vc
  x^*|\bt^*)$ and calculate its redefined distance
  $u^*=G(\rho(\vc x^*, \vc y))$.
  \item
  With probability $\min\left(1,\exp\left[-(u^*-u)/\epsilon^e\right]\right)$
  update $E$, i.e., replace particle $(\bt,u)$ by $(\bt^*,u^*)$.
  \item
    Whenever a significant fraction of the ensemble has been updated,
    update the ensemble average $U$, the transition temperature
    $\epsilon^e$ solving equation (\ref{cooling}) and, optionally, the jump distribution according to eq. (\ref{beta2}).
  \item
   Stop the algorithm if the acceptance rate drops below a certain value.
\end{enumerate}
\end{framed}
\caption{Algorithm I for the case of a non-informative prior.}
\label{AlgI}
\end{table}

The algorithm presented here will not only yield a sample from an
approximation of the posterior, but it will also provide information about
the bias, expressed through the final value of $\epsilon(t)$.
This information, of course, can be used to {\em reduce the bias}, at the
cost of sacrificing some effective sample size, via attaching the weights
$\exp(-\delta u(\vc z)/\epsilon)$, with $\delta$ being a small
dimensionless parameter, to the final ensemble and re-sampling a new
ensemble according to these weights.
The choice of $\delta$ is arbitrary and expresses the trade-off between
bias and effective sample size of the ensemble.
The weights were chosen such that the re-sampled ensemble still represents
a distribution of the form (\ref{muapprox}). Thus, such a bias correction
step can also be applied, occasionally, during the algorithm, as long as
the ensemble is given enough time to recover from the loss of effective
sample size between two resampling steps.

Let us conclude this subsection with comments  on the adaptive choice of
the covariance $K$ of the jump distribution $k$.
To this end, we choose, for $k(\bt,\bt')$, a symmetric normal jump
distribution, whose covariance is adapted to the empirical covariance of
the marginal of $\mu(\vc z,t)$, $\Sigma(t)$, according to eq.
\begin{equation}
\label{beta2}
  K=\beta\Sigma(t)+s\tr(\Sigma)\identy\,
\end{equation}
where $s$ is a small constant preventing (\ref{beta2}) from degenerating
and $\beta$ is an additional tuning parameter of the algorithm that mustn't
be chosen much smaller than unity in order that the mixing in parameter
space is fast enough compared to the decay of the mean distance to the
target.
Note that our derivation of the optimal cooling schedule was based on the
assumption of a time-constant $k(\bt,\bt')$.
The adaptation (\ref{beta2}) makes $k(\bt,\bt')$ time-dependent, which
leads to two compensatory effects.
On the one hand, due to the increased acceptance probability ensued by this
adaptation, the optimal schedule would be given by a time-dependent tuning
parameter $v(t)$ that increases with time. This can be seen by repeating
the exercise in \cite{spirkl_1995_optFiniteTimeEndorevProc}, with an
explicitly time-dependent $\dot U=\dot U(U,F,t)$, and acknowledging the
fact that $\partial\dot U/\partial t<0$ if the adaptation (\ref{beta2})
leads to an increase of the acceptance rate, relative to a schedule without
adaptation.
On the other hand, typically, adaptation makes $k(\bt,\bt')$ sharper over time and, therefore, $\gamma$ tends to increase over time.
Thus, if we set $v/\gamma=\const$, $v$ tends to increase over time.
In general, the optimal schedule for $\epsilon^e(t)$, if adaptation
(\ref{beta2}) is employed, cannot be determined easily.
Therefore, the best strategy seems to be to turn on adaptation
(\ref{beta2}) and check whether the gain of efficiency due to an increased
acceptance rate offsets the loss due to the deviation from the minimal
entropy production path.

At this time, it would be premature to come up with too many
recommendations of how to choose the tuning parameters $v$ and $\beta$, as
we do not yet have enough practical experience with the algorithm (but see
the recommendation given in the application part of this paper).
But we want to point out again that a too large $v$ combined with a too
small $\beta$ might lead to a deviation from assumption (\ref{muapprox})
and, therefore, a bias that would be impossible to correct for.


\subsubsection{The case with an informative prior}\label{SectlinSABC}

As we have discussed at the beginning of this section, the transition rate
(\ref{transition}) has the disadvantage that a too fast decrease of
$\epsilon$ can lead to convergence to a biased result with
under-represented prior.
To account for this bias, and ultimately control it, we replace
(\ref{transition}) by a transition rate with a two-dimensional
 {\em   control parameter} $\pmb \epsilon=(\epsilon_1,\epsilon_2)$,
\begin{equation}\label{newtransition}
q_{{\pmb\epsilon}}((\bt',\vc x'),(\bt,\vc x))=k(\bt',\bt)f(\vc x|\bt)
\min
\left(1,
	\exp\left[
        -\frac{\rho(\vc x)-\rho(\vc x')}{\epsilon_1}
        -(1+\epsilon_2)(\nu(\bt)-\nu(\bt'))
    \right]
\right)\,,
\end{equation}
where
\begin{equation}
  \nu(\bt)=-\ln\left(f(\bt)\right)\,
\end{equation}
and $\rho(\vc x)=\rho(\vc x,\vc y)$.
Transition rate (\ref{newtransition}) satisfies the
{\em detailed balance condition}
\begin{equation}\label{detailedbalance}
  \pi_{{\pmb\epsilon}}(\bt',\vc x')
  q_{{\pmb\epsilon}}((\bt',\vc x'),(\bt,\vc x))
  =
  \pi_{{\pmb\epsilon}}(\bt,\vc x)
  q_{{\pmb\epsilon}}((\bt,\vc x),(\bt',\vc x'))\,,
\end{equation}
for the equilibrium distribution
\begin{equation}\label{newpi}
  \pi_{{\pmb\epsilon}}(\bt,\vc x)
  =
  Z^{-1}({\pmb\epsilon})
  f(\vc x|\bt)
  e^{-\rho(\vc x)/\epsilon_1-(1+\epsilon_2)\nu(\bt)}\,,
\end{equation}
with
\begin{equation}
  Z({\pmb\epsilon})
  =
  \int
  f(\vc x|\bt)
  e^{-\rho(\vc x)/\epsilon_1-(1+\epsilon_2)\nu(\bt)}
  d\bt d\vc x\,.
\end{equation}

As before, we distinguish between the parameter $\pmb\epsilon^e(t)$
that is used in the transition at time $t$, thus controlling the annealing schedule, and the parameter
$\pmb\epsilon(t)$ which describes the distribution of the process at
time $t$ under the endoreversibility assumption
\begin{equation}\label{muapprox2}
  \mu(\vc z,t)\approx \pi_{{\pmb\epsilon(t)}}(\vc z)\,.
\end{equation}
Again our goal is to find a cooling schedule $\pmb\epsilon^e(t)$
depending on $\pmb\epsilon(t)$ such that the entropy production is
minimized. In addition, we want the prior bias, measured by $\epsilon_2(t)$,
to go to zero.

Initially, at time $t=0$, the distribution is chosen as (\ref{newpi}), with
a rather large $\epsilon_1(0)$ and $\epsilon_2(0)=0$. The corresponding
ensemble is generated by adopting a rejection technique. The first control
parameter $\epsilon^e_1(0)$ is set somewhat smaller than $\epsilon_1(0)$
and  $\epsilon_2^e(0)=0$.

Under the endoreversibility assumption, the distribution at any time
is now characterized by the following two expectations (``extensive
thermodynamic quantities'')
\begin{align}
  U_1(t)&:=\int \rho(\vc x)\mu(\bt, \vc x,t)d\bt d\vc x\,,\label{U1}\\
  U_2(t)&:=\int \nu (\bt  )\mu(\bt, \vc x,t)d\bt d\vc x\,.\label{U2}
\end{align}
By standard results about exponential families, there is a one-to-one
correspondence between the vectors $\vc U$ and the parameters
(intensive quantities) ${\pmb\epsilon}$. This allows us to describe the
system by the time-dependent vector ${\pmb\epsilon}(t)={\pmb\epsilon}(\vc
U(t))$. We are however not able to achieve approximate equality
of these two vectors by a simple transformation.

As in the previous subsection, the entropy production rate can be
expressed as
$$
\dot S_{irr}=\vc F(t)^T\dot{\vc U}(t)\,,
$$
where the driving forces are now
$$
\vc F(t)
=
\begin{pmatrix}
\epsilon_1(t)^{-1} - \epsilon^e_1(t)^ {-1}\\
\epsilon_2(t) - \epsilon^e_2(t)\,
\end{pmatrix}\,.
$$
In order to find a necessary condition for minimal entropy production,
we need as before to express $\vc F(t)$ as a function of $\vc U(t)$ and
$\dot{\vc U}(t)$ and to compute in particular the matrix of partial
derivatives $\frac{\partial \vc F}{\partial \dot{\vc U}}$.

In this two-dimensional setting, it seems however infeasible to establish
a  non-linear relationship between $\vc F$ and $\dot{\vc U}$ as we did in
(\ref{Udot10}) for the one-dimensional setting.
Therefore, we shall make the {\em linearity assumption}
\begin{equation}\label{Onsager}
  \dot{\vc U} \approx L(\vc U)\vc F\,,
\end{equation}
which is reasonable as long as $\vc F(t)$ is not too large.
Using the detailed balance condition (\ref{detailedbalance}), we find
\begin{multline}\label{L}
  L_{ij}(\vc U)=Z^{-1}({\pmb\epsilon})
  \int
  (u_i(\vc z)-u_i(\vc z'))
  (u_j(\vc z)-u_j(\vc z'))
  k(\bt,\bt')\\
  \times f(\vc x|\bt)f(\vc x'|\bt')
  \exp[
    -\rho(\vc x)/\epsilon_1-(1+\epsilon_2)\nu(\bt)
    ]\\
  \times\chi\left(
    (\rho(\vc x)-\rho(\vc x'))/\epsilon_1+(1+\epsilon_2)(\nu(\bt)-\nu(\bt'))
    \right )
  d\vc xd\vc x'd\bt d\bt'\,,
\end{multline}
with $u_1(\vc z)=\rho(\vc x)$ and $u_2(\vc z)=\nu(\bt)$. The $\vc U$ dependence of the r.h.s. of (\ref{L}) is through ${\pmb\epsilon}={\pmb\epsilon}(\vc U)$.
The matrix $L$ is {\em symmetric} and {\em positive definite} (due to the
Cauchy-Schwarz inequality).
In the theory of non-equilibrium thermodynamics, the entries of the matrix
$L$ are known as the {\em Onsager coefficients}
\cite{onsager_1931_IrreversibleProcessesI}.

In two dimensions, eq. (\ref{preschedule}) becomes the necessary condition for minimal entropy production
\begin{equation}\label{preschedule2}
 \dot{\vc U} ^T \frac{\partial \vc F}{\partial \dot{\vc U}}
 \dot{\vc U} = \const = v\,.
\end{equation}
Plugging (\ref{Onsager}) into (\ref{preschedule2}) we find a necessary
criterion for optimality to be given by
\begin{equation}\label{constentropyprod}
  \dot{\vc U} ^T R(\vc U)
 \dot{\vc U} = v\,,
\end{equation}
where $R(\vc U):=L^{-1}(\vc U)$ defines a metric on the $(U_1,U_2)$-plane.
Equation (\ref{constentropyprod}) can also be derived as follows:
Under the linearity assumption (\ref{Onsager}), and due to the Cauchy-Schwarz inequality, the entropy production satisfies the inequality
\begin{equation}\label{sigma}
  S_{irr} =\int_{0}^{t_f}
  \dot{\vc U}(t) ^T R(\vc U(t)) \dot{\vc U}(t) dt
  \geq
  \frac{\mathcal K}{t_f}\,,
\end{equation}
where $\mathcal K$ is the length of the process-path in the $(U_1,U_2)$-plane, measured with the metric $R(\vc U)$.
The lower bound of (\ref{sigma}) is assumed if the integrand is constant,
i.e., if the entropy production rate is constant
\cite{salamon_1980_optHeatEngine}.
Thus, finding the optimal schedule consists in (i) finding the shortest
path in the $(U_1,U_2)$-plane and (ii) traveling along this path such that
the entropy production rate is constant.
Therefore, condition (\ref{constentropyprod}) completely determines the
optimal trajectory, which is of course a consequence of the linearity
assumption (\ref{Onsager}).

In order to define our algorithm, we  have to continuously estimate
the following quantities during run-time: (i) the ensemble means $\vc
U(t)$,  (ii) the intensities ${\pmb\epsilon}(t) = {\pmb\epsilon}(\vc U(t))$ that
determine our system under assumption (\ref{muapprox2}) and (iii) the
metric $L(\vc U(t))$. As $(i)$ is trivial, we now discuss (ii) and (iii).

Given a small change, $\Delta\vc U$, of the ensemble means, the corresponding change of the intensities, $\Delta{\pmb\epsilon}$, is estimated by means of
\begin{equation}
\label{Delta-eps}
\Delta{\pmb\epsilon}
\approx
\left(
    \frac{\partial\vc U}{\partial{\pmb\epsilon}}
\right)^{-1}
\Delta\vc U\,,
\end{equation}
where the Jacobi matrix
\begin{equation}\label{Jacobi}
      \frac{\partial \vc U}{\partial{\pmb\epsilon}}:=
      \begin{pmatrix}
       \frac{1}{\epsilon_1^2}\Var(\rho)
            &   -\Cov(\rho,\nu) \\
       \frac{1}{\epsilon_1^2}\Cov(\rho,\nu)
            &   -\Var(\nu)
     \end{pmatrix}
\end{equation}
is estimated using the empirical covariance matrix of the $\rho$ and $\nu$ components of the ensemble.
However, the neglected higher order corrections will eventually lead to large deviations from the "true" state.
Therefore, occasional corrections have to take place estimating $\vc U({\pmb\epsilon})$ without using the ensemble $E$.
Such an estimate can be calculated using the ensemble $P$ drawn initially from the joint prior $f(\vc x,\bt)$.
Once ${\pmb\epsilon}$ is estimated, we need to estimate $L(\vc U)$ in order to determine the adaptive tuning parameters ${\pmb\epsilon}^e$.
Inspecting equation (\ref{L}) reveals that this can be done using the prior sample $P$ as well as the ensemble $E$.
This estimate relies on assumption (\ref{muapprox2}).
At the end of this subsection we will discuss a way of improving both estimates, $\vc U({\pmb\epsilon})$ and $L(\vc U)$, for small $\epsilon_1$, when the effective sample size of $P$ is low.

Since, at the beginning of the algorithm, neither is the target value for $U_2$, at $\epsilon_1=\epsilon_2=0$, known exactly nor is the metric $R(\vc U)=L^{-1}(\vc U)$ known globally. Therefore, it appears difficult to come up with an optimal path in the $(U_1,U_2)$-plane.
However, it appears reasonable to force the process to be on a path such that $\epsilon_2$ remains small.
Practically, this can be achieved by applying a counter force, setting
\begin{equation}
  \epsilon_2^e=-a\epsilon_2\,,
\end{equation}
where $a$ is some positive constant.
Finally, in order to find the optimal trajectory, under these restrictions, we need to choose
$\epsilon_1^e$ such that (\ref{constentropyprod}) is satisfied. Using
(\ref{Onsager}), we obtain the quadratic equation
\begin{equation}\label{constentropyprodL}
  \vc F^T L(\vc U) \vc F=v\,.
\end{equation}
The easiest version of the algorithm presented in this subsection is summarized in Table \ref{AlgII}.

Note that the prior-bias in the final ensemble, as expressed through $\epsilon_2(t)$, can be completely corrected via a weighted re-sampling, in much the same way as the bias due to a non-vanishing $\epsilon_1(t)$ was reduced in the last subsection.

In the remainder of this subsection, we outline two alternative ways of estimating ${\pmb\epsilon}(\vc U)$ and $L(\vc U)$, using the information gathered during the course of the algorithm. They can be used when $\epsilon_1$ gets very small and the prior sample $P$ yields poor estimates.
Both methods, however, will depend on the assumption (\ref{muapprox2}) being satisfied.
One way is to simply correct the ensemble $E$ with weights proportional to $e^{-\rho(\vc x)/\epsilon_1-\epsilon_2\nu(\bt)}$, in order to get a new prior sample, which has a better resolution where $\epsilon_1$ is small.
The other way is to populate, during the course of the algorithm, a {\em transition matrix}, $Q$, of {\em attempted} moves \cite{andresen_1988_lumpedThdynPropSA}.
That is, we partition an area of interest in the $(U_1,U_2)$-plane (which will contain the small distances $\rho$) into $n_{U_1} n_{U_2}$ bins and increment the matrix element $Q^{ij}{ }_{i'j'}$, whenever a particle in bin $U_{1,i'}\times U_{2,j'}$ attempts to move into bin $U_{1,i}\times U_{2,j}$.
In order to get the correct transition matrix the diagonal entries $Q^{i'j'}{}_{i'j'}$ must be incremented whenever a particle from bin $U_{1,i'}\times U_{2,j'}$ attempts to jump outside the area of interest. Furthermore, the columns of $Q$ must be normalized so that their sums equals unity.
Under assumption (\ref{muapprox2}) it holds that
$$
Q^{ij}{ }_{i'j'}
=
\frac
{\int_{\rho(\vc x)\in U_{1,i}\,,
    \rho(\vc x')\in U_{1,i'}\,,
    \nu(\bt)\in U_{2,j}\,,
    \nu(\bt')\in U_{2,j'}}
k(\bt,\bt')f(\vc x|\bt)f(\vc x'|\bt')
d\vc xd\vc x' d\bt d\bt'
}
{\int_
    {\rho(\vc x')\in U_{1,i'}\,,
    \nu(\bt')\in U_{2,j'}}
f(\vc x'|\bt')
d\vc x' d\bt'}\,.
$$
The eigenvector, $\vc g$, corresponding to the largest eigenvalue, 1, of $Q$, is a discretization of the likelihood function on the $(U_1,U_2)$-plane:
\begin{equation}\label{g}
g_{i'j'}
=
\frac
{\int_
    {\rho(\vc x')\in U_{1,i'}\,,
    \rho(\bt')\in U_{2,j'}}
f(\vc x'|\bt')
d\vc x' d\bt'}
{\int f(\vc x|\bt)d\vc xd\bt}\,.
\end{equation}
This holds true even if $k$ is adapted during the algorithm.
At a later stage of the algorithm, when the prior sample becomes insufficient but $Q$ is sufficiently well populated to estimate (\ref{g}), the latter can be used to estimate both $\vc U({\pmb\epsilon})$ and $L(\vc U)$, for small values of $\epsilon_1$.
Furthermore, if $k$ is not adapted, the matrix $Q$ can be used directly to estimate $L(\vc U)$, without the need of calculating the jump density matrix $K$.

Of course, a similar matrix of attempted moves can also be used in the one-dimensional setting, subsection \ref{nlSABC}, to replace prior sample $P$ at a later stage of the algorithm.

\begin{table}
\vspace{-2cm}
\begin{framed}
{\bf Input:}
\begin{enumerate}
\itemsep-0.4em
   \item Algorithms to sample from the prior and the likelihood.
   \item Ensemble size $N$ and initial value $\epsilon_{init}$.
   \item Tuning parameters $\beta$, $s$, $v$ and $a$ with default values,
   $\beta=2$, $s=0.01$, $v=0.3$ and $a=2$.
\end{enumerate}
{\bf Initialization:}
\begin{enumerate}
\itemsep-0.4em
  \item
  Repeat, until the ensemble $E$ in (d) contains $N$ particles:
  \begin{enumerate}
  \itemsep-0.4em
    \item
    Sample a parameter vector, $\bt$, from the prior.
    \item
    Sample an output, $\vc x$, from the likelihood $f(\vc x|\bt)$.
    \item
    Store the vector $(\bt, \rho(\vc x,\vc y),v(\bt))$ in ensemble $P$.
    \item
    With probability $\exp[-\rho(\vc x,\vc y)/\epsilon_{init}]$ store the vector $(\bt,\rho(\vc x,\vc y),\nu(\bt))$ also in ensemble $E$.
  \end{enumerate}
  \item Initialize metric $L(\vc U)$ defined in (\ref{L}), using the prior
          ensemble $P$.
  \item
  Initialize $\vc U$ as the ensemble $E$ averages.
    \item
  Initialize $\epsilon_2^e=0$ and $\epsilon_1^e$ solving the quadratic eq. (\ref{constentropyprodL}).
  \item
  Initialize $K$ according to (\ref{beta2}).
\end{enumerate}

{\bf Iteration:}
\begin{enumerate}
\itemsep-0.4em
  \item
  Select an arbitrary particle, $(\bt,\rho,\nu)$, from the ensemble $E$.
  \item
  Sample a proposal parameter vector, $\bt^*$, from $k(\bt,\bt^*)$ and
  a proposal output, $\vc x^*$, from the likelihood $f(\vc x^*|\bt^*)$.
  \item
  With probability $r=\min\left(
  1,
  \exp\left[
            -(\rho^*-\rho)/\epsilon^e_1
            -(1+\epsilon^e_2)(\nu^*-\nu)
        \right]
  \right)$,
    update $E$, i.e., replace $(\bt,\rho,\nu)$ by $(\bt^*,\rho^*,\nu^*)$.
  \item
    Whenever a significant fraction of the ensemble has been updated,
    perform the following mean-field updates:
    \begin{itemize}
    \itemsep-0.4em
      \item
      Save the old ensemble means $\vc U_{old}$ and denote the new ones by
      $\vc U_{new}$.
      \item
      Update the Jacobi matrix (\ref{Jacobi}) via calculation of the
      empirical covariance matrix of the $\rho$ and $\nu$ components of
      $E$.
     \item
     Save the old intensities ${\pmb\epsilon}_{old}$ and calculate the new
     ones iterating the following two steps:
     \begin{enumerate}
      \item
      Compute the change
      $\Delta {\pmb\epsilon}= {\pmb\epsilon}_{new}-{\pmb\epsilon}_{old}$
      according to equation (\ref{Delta-eps}).
      \item
      If $\vc U$ is close (say within a relative error of $1\%$) to the
      theoretical ensemble averages, $\vc U({\pmb\epsilon}_{new})$,  as
      calculated from $P$, set ${\pmb\epsilon}={\pmb\epsilon}_{new}$ and
      stop, otherwise, replace ${\pmb\epsilon}_{new}\rightarrow
      {\pmb\epsilon}_{old}$ and $\vc U({\pmb\epsilon}_{new})\rightarrow \vc
      U_{old}$ and go back to (a).
    \end{enumerate}
    \item
    Update the metric $L(\vc U)$, according to ${\pmb\epsilon}$, using
    prior ensemble $P$.
     \item
      Update $\epsilon^e_2=-a\epsilon_2$ and $\epsilon^e_1$ solving (\ref{constentropyprodL}).
      \item
      Optionally: update $K$ according to (\ref{beta2}).
    \end{itemize}
    \item
   Stop the algorithm if the acceptance rate drops below a certain value.
\end{enumerate}
\end{framed}
\caption{Algorithm II, for the case of an informative prior.}
\label{AlgII}
\end{table}

\section{Toy examples}\label{Toys}

In this section, we apply our adaptive scheme to two examples. The
prior of the first one has almost no influence on the posterior, in
the second this influence is large.
As a shorthand for our adaptive scheme we use the acronym SABC, which merges SA, for Simulated Annealing, with ABC.

SABC is compared against the sequential Monte Carlo samplers (SMC)
from del Moral et al.\ (2012) \cite{del2012adaptive} and adaptive
population Monte Carlo (APMC) from Lenormand et al.\ (2013)
\cite{Lenormand_2012_ABC}. For the latter two the implementations in
the \textsf{R}-package ``EasyABC'' \cite{Jabot_2013_ABC} were used.

For SMC and APMC the same tuning parameters were used for both
examples. The population size $N$ for all algorithms was 1000.
The parameter $\alpha$ of APMC was set to $0.5$ following the recommendation of
Lenormand et al.\ (2013).  The tuning parameters for SMC are the same del Moral et al.\ (2012) used for the first toy example
($\alpha=0.95$, $M=1$, $N_T=500$).

In real applications the computational costs are often dominated by
sampling from the likelihood. Therefore, the number of samples drawn
from the likelihood was used as measure of the computational effort.

\subsection{Example 1}

The first example is a traditional example of the ABC literature
(e.g.\ \cite{del2012adaptive},
\cite{Lenormand_2012_ABC}).  The prior is uniformly distributed on the
interval $[-10,10]$, and the likelihood is given by the sum of two
normal distributions with very different standard deviations:
\begin{equation}
  f(x|\theta)
  \propto
  \exp\left[
    -\frac{(x-\theta)^2}{2}
    \right]
    +
    \frac{1}{\sigma}\exp\left[
        -\frac{(x-\theta)^2}{2\sigma^2}
    \right]\,,
\end{equation}
with $\sigma=0.1$.
Thus, the posterior for $y=0$ is given by
\begin{equation}
  f(\theta|y) \propto
  \identy_{[-10,10]}
  \left(
    \exp\left[
        -\frac{\theta^2}{2}
    \right]+
    \frac{1}{\sigma}\exp\left[
        -\frac{\theta^2}{2\sigma^2}
    \right]
  \right)\,.
\end{equation}

As the prior has almost no influence on the posterior, the non-linear algorithm from Table \ref{AlgI}, with one final bias correction, has been employed.
Furthermore, $k(\bt,\bt')$ has been continuously adapted according to (\ref{beta2}).
Since the prior has a much bigger variance than the posterior, this turns out to be beneficial, for the convergence of the algorithm.
The optimal choice for the dimensionless parameter $\beta$, defined in (\ref{beta2}), is expected to depend little on details of the model. For our examples we choose $\beta=2$, which is large enough to ensure a fast enough mixing in parameter space compared to the decay of the mean distance to the target.
The tuning parameter $v/\gamma$ governs the annealing speed.
As this example is so simple, its choice is not very critical. With the choice $\beta=2$ a violation of the endoreversibility assumption is not to be expected, even for high annealing speeds. A slowing down of the convergence due to a trapping of particles is observed only at very high values of $v/\gamma$.
We choose the value $v/\gamma = 3$.

\begin{figure}[h]
  \includegraphics[width=14cm]{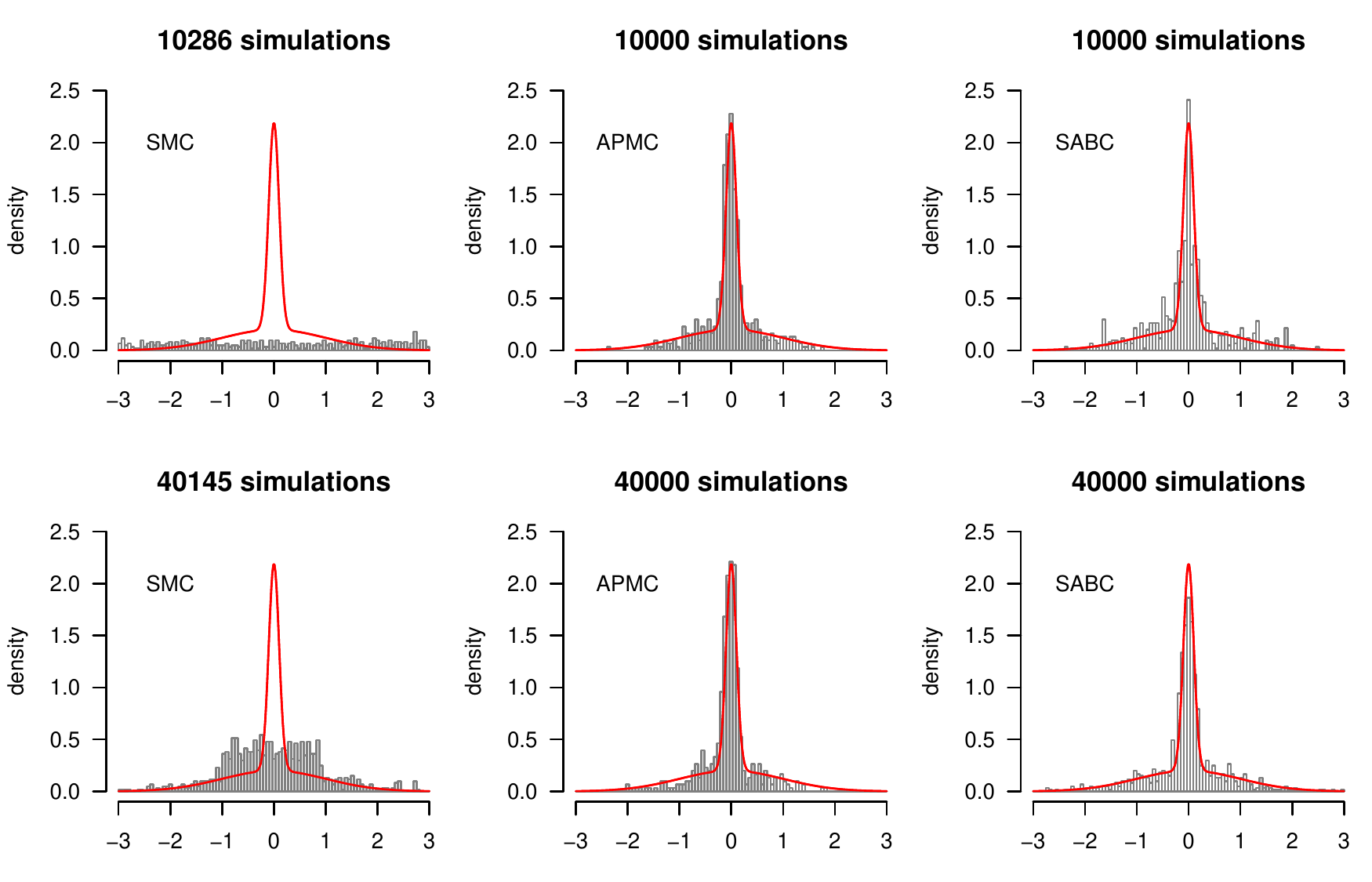}
  \caption{Histograms of an ensemble of 1000 particles for example~1
    generated with SMC, APMC and SABC. The solid curve is the exact posterior density.
    Note that "simulations" refers to single draws from the likelihood.}\label{FigToy1}
\end{figure}

Figure~\ref{FigToy1} shows the results for all three samplers after, approximately,
10\,000 and 40\,000 simulations from the likelihood. It is clearly visible that
SMC has not yet converged, while the results of APMC and SABC look
much better. After 40\,000 likelihood
samples, the histogram of SABC looks  slightly smoother than the one of APMC.
As APMC is an importance sampling algorithm, the sample generated after 40\,000 simulations is an exact sample from a closer approximation of the posterior than the sample generated after 10\,000 simulations. Therefore, we attribute the slight deterioration of the histogram to the loss of effective sample size (ESS) due to resampling.
The ESS, for APMC and SABC, are summarized in Table \ref{TableESS}.
For APMC, the ESS was calculated under the optimistic assumption that, before the last resampling is made, the ensemble has completely recovered from the loss of ESS.
For SABC, the loss of ESS after 10000 simulations is due to the final bias correction step. The parameter $\delta$, used for the final bias correction as described towards the end of subsection \ref{nlSABC}, was chosen such that the ESS of SABC and APMC are comparable.

\begin{table}[h]
\centering
\begin{tabular}{c|c|c}

			&	APMC	&	SABC	\\
\hline
10\,000 simulations	&	306		&	240		\\
40\,000 simulations   &  323        & 1000

\end{tabular}
\caption{Comparison of effective sample sizes of the APMC and the
  SABC algorithm for example~1, after 10\,000 and 40\,000 likelihood simulations.}
\label{TableESS}
\end{table}

\subsection{Example 2}

In contrast to the first example, the prior in the second example
has a large influence on the posterior.
The prior shall be given as the normal distribution
$$
f(\theta)=\frac{1}{\sqrt{2\pi}}\exp\left[-\frac{\theta^2}{2}\right]\,,
$$
and the likelihood as the normal distribution
$$
f(x|\theta)=\frac{1}{\sqrt{2\pi}}\exp\left[-\frac{(x-\theta)^2}{2}\right]\,.
$$
Thus, the posterior is given as
$$
f(\theta|y)=\frac{1}{\sqrt{\pi}}\exp\left[-{(\theta-y/2)^2}\right]\,.
$$
To investigate if the algorithms can handle severe {\em prior-data conflicts}, we set $y=3$.

In this example it is important for SABC to properly control
$\epsilon_2(t)$ while annealing $\epsilon_1(t)$ as prior and
likelihood "pull from opposite directions".
Therefore, we employ the linear algorithm as described in Table \ref{AlgII}, with one final bias correction.
The tuning parameter $v$, which now has the interpretation of an entropy production rate, was chosen to be $0.3$. For $\beta$ we chose the same value as in the previous example, namely $\beta=2$.
In this example, continuously adapting $k(\bt,\bt')$ has a negligible effect on the convergence speed.

\begin{figure}[h]
  \includegraphics[width=14cm]{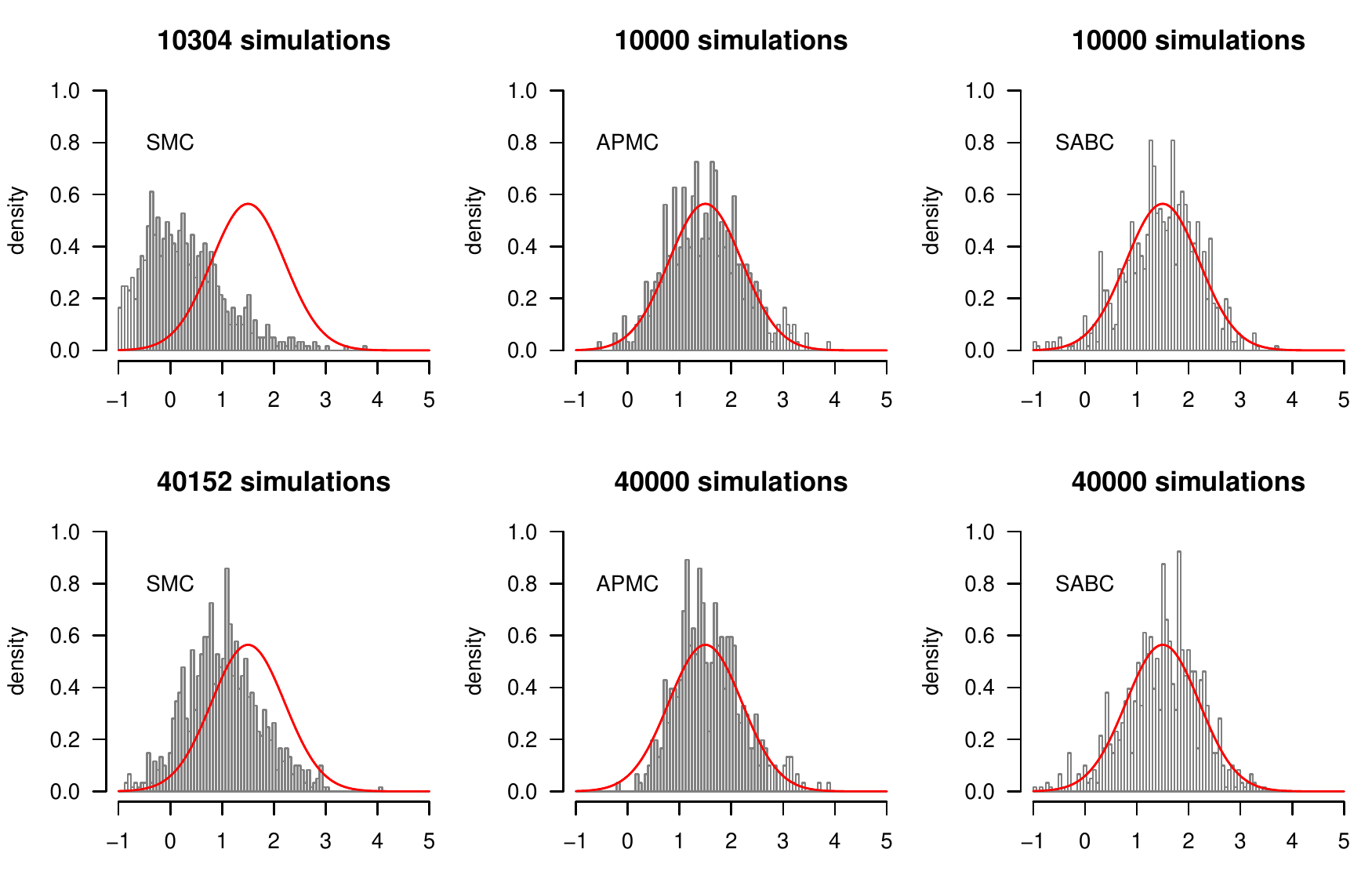}
  \caption{Histograms of an ensemble of 1000 particles for example~2
    generated with SMC, APMC and SABC. The solid curve is the exact posterior density.}\label{FigToy2}
\end{figure}

The results are shown in figure \ref{FigToy2}. Again, the results from
SMC have not yet converged and are heavily biased towards the
prior. APMC seems to converge slightly faster then SABC (compared
 at 10\,000 simulations). However, the quality of the APMC sample decreases
 for more simulations, which is attributed to the loss of ESS. As SABC is avoiding resampling,
 this effect is not observed.
Effective sample sizes, for APMC and SABC are summarized in Table \ref{TableESS2}.
After 10000 simulations, we chose the bias-correcting parameter $\delta$ such that the ESS of SABC and APMC are similar. After 40000 simulations, the loss of ESS for SABC is due solely to the correction of the prior bias, expressed through the final value of $\epsilon_2$.

\begin{table}[h]
\centering
\begin{tabular}{c|c|c}

			&	APMC	&	SABC	\\
\hline
10\,000 simulations	&	404		&	408		\\
40\,000 simulations   &  322        &  982

\end{tabular}
\caption{Comparison of effective sample sizes of the APMC and the
  SABC algorithm for example~2, after 10\,000 and 40\,000 likelihood simulations.}
\label{TableESS2}
\end{table}

\section{Real-world example: tuberculosis bacteria}\label{CaseStudy}

Tanaka et al. \cite{tanaka_2006_ABC} analyzed genotype data of tuberculosis bacteria
with a stochastic model to infer death, birth and
mutation rates by means of ABC. In the 473 analyzed tuberculosis bacteria cultures, 326
distinct genotypes where found. Cultures with the same genotype form
a cluster. The data in table~\ref{tab:tuber} describe how many
clusters with a certain number of cultures were found. For example
one cluster consisting of 30 cultures with the same genotype was observed, two
clusters with five cultures each, and so forth.

This data contains only information on the rates relative to each
other, because no time information is available. Therefore, only
birth, death, and mutation events are simulated until the population
reaches a (arbitrarily defined) size of 10\,000 living bacteria (see \cite{tanaka_2006_ABC} for details). Therefrom a random sample without replacement of size
473 is taken. We used the parametrization proposed by Fearnhead and
Prangle \cite{Fearnhead_2012_ABC} which reduces the inference to a two dimensional
problem with $a = P(\text{birth}|\text{event})$ and $d =
P(\text{death}|\text{event})$. The probability that an event is a
mutation is given by $1-a-d$. Also the same flat prior is used $\pi(a,
d) \propto \mathbf{1}_{a>d} \mathbf{1}_{0<a+d\leq1}$.

\begin{table}[b]
  \centering
  \caption{Tuberculosis genotype data. }
  \begin{tabular}{lrrrrrrrrrr}
      \hline
      number of cultures per cluster & 30 & 23 & 15 & 10 & 8 & 5 &  4 &  3 & 2 & 1 \\
      number of clusters &  1 &  1 &  1 &  1 & 1 & 2 & 4 & 13 & 20 & 282 \\
      \hline
  \end{tabular}
  \label{tab:tuber}
\end{table}

The data are summarized by two statistics as described by \cite{tanaka_2006_ABC}: the number of distinct genotypes $g$ in the sample and a
measure of gene diversity $H=1-\sum_{i=1}^g \left(n_i/473 \right) ^2$,
where $n_i$ is the number of bacteria in the $i$th cluster. The
distance between simulated and observed data is measured as
$|g^*-g|/473 + |H^*-H|$, where the asterisks indicate the statistics of
the simulated data.

Because of the flat prior the non-linear version of the SABC, Table \ref{AlgI}, was
used, with a final re-sampling step, with $\delta=0.2$.
As in the previous examples, we chose $\beta=2$ and tuned $v/\gamma$. A high convergence speed is achieved with $v/\gamma=7$ but we found that the algorithm is remarkably robust w.r.t. the choice of this tuning parameter.

\begin{figure}[h]
\centering
  \includegraphics[width=14cm]{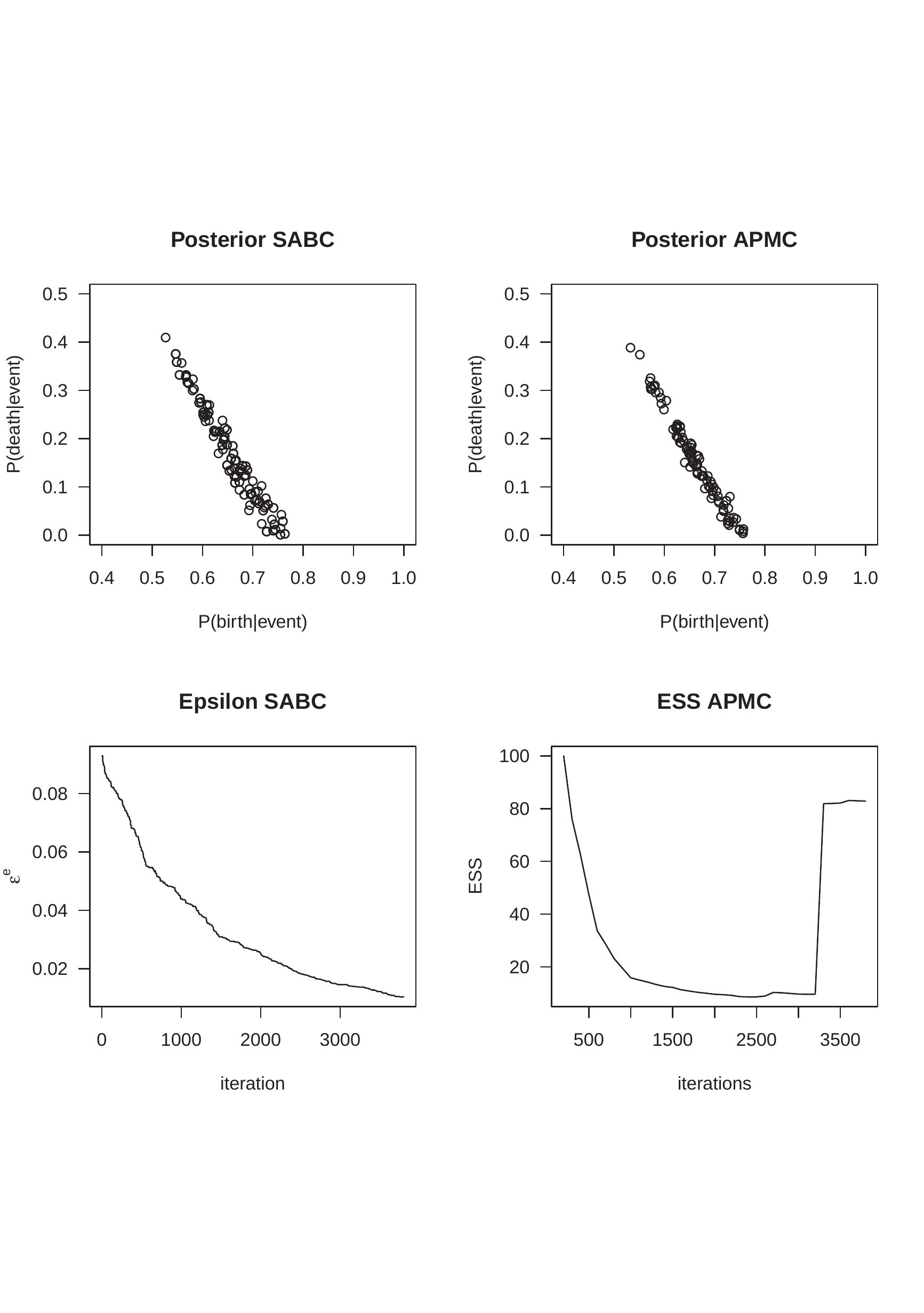}
  \caption{Top row: Final population of 200 particles after a total of 3800 updates (approximately 2000 simulations from the likelihood, the rest were jumps into forbidden parameter regions), for SABC (left) and APMC (right).
  Bottom row: Time-course of $\epsilon^e(t)$, for SABC (left) and time-course of the ESS, for APMC (right). The final ESS, for SABC, is 129 and due to a single resampling step.}\label{FigReal}
\end{figure}

We compared the performance of our algorithm with the adaptive
population Monte Carlo (APMC) from Lenormand et al. \cite{Lenormand_2012_ABC}, which we ran with the same sample size $N=200$ and with the choice of the tuning parameter $\alpha=0.5$, as recommended in \cite{Lenormand_2012_ABC}.
Similarly to the results from the previous section, we found that, for short simulation times, APMC shows a slightly better performance than SABC due to a faster convergence, but, for longer simulation times, APMC suffers from a deterioration of the sample due to a loss of ESS, which is not observed with SABC.
Figure \ref{FigReal} shows the results after 3800 iterations (approximately 2000 simulations from the likelihood).
The result of SABC is in excellent agreement with the result reported in \cite{Fearnhead_2012_ABC}, whereas the final sample from APMC shows some signs of deterioration, which is attributed to the loss of ESS, in each iteration step.
The time-course of the ESS, for APMC, is shown in Figure \ref{FigReal}. The jump to an ESS of about 80, before the last resampling step, is based on the (presumably unrealistic) assumption that each population update leads to a complete recovery of the ESS.
For SABC, the final ESS is 129 and due to the final resampling step.

\section{Conclusions}\label{SectConclusions}

We have presented a framework of particle algorithms for Approximate Bayes Computations that is inspired by Simulated Annealing.
Its main advantage compared to the sequential ABC algorithms the authors are aware of is the fact that it is not based on importance sampling.
Therefore, the effective sample size of our algorithms does not decrease over time.
As the interactions between the particles in the adaptive algorithm are of {\em mean-field type}, the statistical independence of the particles is preserved (see, e.g., \cite{burkholder_1991_PropagationOfChaos}).

The cost for this gain of efficiency is the fact that our system is necessarily out of equilibrium. That is, in addition to the bias due to non-zero equilibrium tolerances $\epsilon_1^e$ and $\epsilon_2^e$, we have a bias due to our system being out of equilibrium (i.e. $\epsilon_1$ being larger than $\epsilon_1^e$).
There is a trade-off between these two kinds of bias reflected in the choice of the tuning parameter $v$.
Choosing a larger $v$ might result in a smaller $\epsilon_1^e$, for a given computation time, but in a larger bias of the second kind.
Choosing $v$ too large, in combination with too slow mixing in parameter space, expressed through a too small $\beta$, might lead to a third kind of bias, a violation of the endoreversibility assumption (\ref{muapprox}) or (\ref{muapprox2}). This kind of bias is impossible to correct for and has to be avoided by a careful choice of tuning parameters.

In Sect. \ref{Sectexplicit} we proved convergence to the correct posterior, for cooling that is slower than a certain inverse power of time. In Sect. \ref{SectCTS} we presented an adaptive cooling scheme that is designed to achieve convergence to the correct posterior with a minimum of computational effort.
Therefore, the control variable $\epsilon_1^e$ is adjusted according to the particles' distance to the target in such a way that the entropy production in the system, which is a measure for the waste of computation, is minimized.
If the prior is important, a second control variable is used to control its influence.
Using this adaptive scheme, tuning essentially reduces to the choice of $\beta$, related to the mixing speed in parameter space, and $v$, related to the annealing speed.


In our scheme the characteristic function $\chi(\epsilon-\rho(\vc x,\vc y))$, which is often used in ABC calculations, is replaced by the Boltzmann factor $\exp(-\rho(\vc x,\vc y)/\epsilon)$.
With this replacement, moves are not only accepted if they end up in an $\epsilon$-ball around the target but they are more likely accepted if they move {\em closer} to the target.

Finally, our algorithm is of the order $\mathcal O(N)$, with some overhead due to occasional mean-field updates needed for the update of the tolerance(s) and the jump distribution.
Importance sampling algorithms are typically of the order $\mathcal O(N^2)$, due to the weighting step, but see the algorithm by del Moral et al \cite{del2012adaptive}, which scales like $\mathcal O(N)$.
However, all the algorithms mentioned in this article scale like $\mathcal O(N)$ with the number of simulations from the likelihood, which is usually the most costly step.
Like all sequential ABC algorithms, our scheme is well suited for {\em parallelization}.

The overhead, in our scheme, is significantly larger if the prior is informative. Furthermore, in this case, we can only derive an optimal schedule for relatively slow annealing (linearity assumption).
For strongly informative priors, a simple ABC rejection algorithm should be considered as an alternative to a sequential schedule.

The biggest disadvantage inherent to all ABC algorithms is that the tolerance leads to a bias that grows with the dimension of the output space $n$.
Therefore, it is important to use {\em summary statistics} to reduce the output dimension or employ {\em local approximations of the likelihood}, for ABC to be useful for problems with large output dimensions (see, e.g., \cite{Fearnhead_2012_ABC} and \cite{leuenberger_2010_ABC}).

Drawing the initial sample for our adaptive algorithm generates, as a side product, a larger sample from the joint prior. In our adaptive scheme we use this prior information, for the redefinition of the metric (\ref{newmetric}) or to estimate the sample average $\vc U({\pmb\epsilon})$ and the metric $L(\vc U)$.
Note that, at the same time, this information can be used to establish appropriate summary statistics, as described in \cite{Fearnhead_2012_ABC}.

\subsection*{Acknowledgements}

The first author is indebted to Bjarne Andresen for valuable comments on the adaptive algorithm.

\bibliographystyle{plain}
\bibliography{refs}

\end{document}